\documentclass[useAMS,usenatbib]{mn2e}
\usepackage{graphics,epsfig,amssymb}
\usepackage[dvipsnames]{color}

\def \be{\begin{equation}}
\def \ee{\end{equation}}
\def \bea{\begin{eqnarray}}
\def \eea{\end{eqnarray}}
\def\etal{{et al.\ }}
\def \ie{{\it i. e.,}}
\def\lsim{\lower.5ex\hbox{$\; \buildrel < \over \sim \;$}}
\def\gsim{\lower.5ex\hbox{$\; \buildrel > \over \sim \;$}}

\title[Simulation of radiation driven wind from disc galaxies]{
Simulation of radiation driven wind from disc galaxies
}
 
\author[I. Chattopadhyay, M. Sharma, B. B. Nath and D. Ryu]
{Indranil Chattopadhyay,$^1$\thanks{indra@aries.res.in}  Mahavir Sharma,$^2$\thanks{mahavir@rri.res.in}
Biman B. Nath,$^2$\thanks{biman@rri.res.in}, Dongsu Ryu$^3$\\
1. ARIES, Manora Peak, Nainital-263 129, Uttarakhand, India\\
2. Raman Research Institute, Sadashiva Nagar, Bangalore 560080, India\\
3. Department of Astronomy and Space Science, Chungnam National University, 
Republic of Korea\\
}

\newcommand{\pd}{\partial}

\newcommand{\rp}{r\prime}
\newcommand{\pp}{\phi\prime}
\begin{document}
 
 
\maketitle 
 
\label{firstpage}
 
\begin{abstract}
We present 2-D hydrodynamic simulation of rotating galactic winds driven 
by radiation. We study the structure and dynamics of 
the cool and/or warm component($T \simeq 10^4$ K) which is mixed with dust. 
We have taken into account the total gravity of a galactic system that
consists of a disc, a bulge and a dark matter halo.
We find that the combined
effect of gravity and radiation pressure from  a realistic disc
drives the gas  away to a distance of $\sim 5$ kpc in $\sim 37$ Myr
for typical
galactic parameters.
The outflow speed
increases rapidly with the disc Eddington parameter $\Gamma_0(=\kappa I/(2 c G 
\Sigma)$) for $\Gamma_0 \ge 1.5$.
We find that the rotation speed of the outflowing gas is 
$\lesssim 100$ km s$^{-1}$. The wind is confined in a cone which mostly 
consist of low angular momentum gas lifted from the central region.
\end{abstract} 
 

\section{Introduction}
Many galaxies are observed to have moving extraplanar gas, 
generally termed as galactic superwinds (see Veilleux et al. 2005 for a recent review).
Initial observations showed the H$\alpha$ emitting gas above the plane of 
M82 (e.g. Lynds \& Sandage 1963). The advent of X-ray astronomy established yet
another phase of galactic outflows, namely the hot plasma, emitting X-rays in 
the temperature range $0.3\hbox{--}2$ keV (Strickland \etal 2004). Also recent observations have 
revealed 
the existence of molecular gas in these outflows (Veilleux et al. 2009, walter \etal 2002).
Earlier observations were limited to local dwarf starburst galaxies 
that showed these 
winds. However, in recent years, the
observations
of outflows in Ultra Luminous Infra-red Galaxies (ULIGs) have extended the
range
of galaxies in which outflows are found (Martin 2005, Rupke et al. 2005, Rupke et al.
2002). 


On the theoretical side, there have been speculations on 
winds from starburst galaxies 
(Burke 1968, Mathews \& Baker 1971, Johnson \& Axford 1971). In these models the 
large scale winds are a consequence of energy injection
by multiple supernovae (Larson 1974, Chevalier \& Clegg 1985, Dekel \& Silk 
1986, Heckman 2002).
In the context of the multiphase structure of the outflows, the results of these
theoretical models
are more relevant for the X-ray emitting hot wind.
On the other hand,  observations of the cold outflows are better explaind by the radiation driving (Murray \etal 2005, Martin 2005).

If only Thompson scattering is considered, then radiation from galaxies 
does not seem to be a reasonable wind driving candidate because 
opacities would be small; however one should consider that these winds are 
heavily enriched. Murray et al. 2005 proposed
a wind driving mechanism based on the scattering of dust-grains by the photons 
from the galaxy (see also Chiao \& Wickramasinghe 1972; Davies et al. 1998). 
This mechanism can be quite
effective since the opacities in dust-photon scattering can be of the order of 
hundred cm$^2$g$^{-1}$ and
gas in turn, being coupled with the dust, is driven out of the galaxy if the 
galaxy posseses a certain critical luminosity.  Bianchi \& Ferrara (2005)
argued that dust grains ejected from galaxies by radiation pressure 
can enrich the 
intergalactic medium. Nath \& Silk (2009) then described a model
of outflows with radiation and thermal pressure, in the context of outflows from Lyman break galaxies
observed by Shapely \etal (2005). Murray
\etal (2010) have also described a similar model in which radiation pressure is important for the
first few million years of the starburst phase, after which SN heated hot gas pushes the outflowing
material. 
Sharma \& Nath (2011) have also shown that radiation pressure is important for
outflows from high mass galaxies with a large SFR (with $v_c \ge
200$ km s$^{-1}$, SFR $\ge 100$ M$_{\odot}$ yr$^{-1}$),  particularly in ULIGs.

In this paper, we study the effect of radiation pressure in driving 
 cold and/or warm gas outflows
from disc galaxies with numerical simulations. 
Recently, Sharma \etal (2011) calculated the
terminal speed of such a flow along the pole of a disc galaxy, taking into account the gravity of 
disc, stellar bulge and dark matter halo. They determined the minimum luminosity (or, equivalently,
the maximum mass-to-light ratio of the disc) to drive a wind, and also showed that the terminal
speed lies in the range of $2\hbox{--}4 \, V_c$ (where $V_c$ is the rotation speed of the disc
galaxy), consistent with observations (Rupke \etal 2005, Martin 2005), and the ansatz used by numerical simulations
in order to explain the metal enrichment of the IGM (Oppenheimer \etal 2006). We
investigate further the physical processes for a radiation driven wind. 
Rotation is yet another aspect of the winds that we address in our 
simulation. 
As the wind material is lifted from a rotating disc,
it should be rotating inherently which is seen in observations as well 
(Greve 2004, Westmoquette et al. 2009, Sofue et al. 1992, Seaquist \& Clark 2001,
Walter et al. 2002).

Previous simulations of galactic outflows have considered the driving force of 
a hot ISM energized by the effects of supernovae (Kohji \& Ikeuchi 1988; Tomisaka \& Bregman 1993;
Mac Low \& Ferrara 1999; Suchkov 
\etal 1994, 1996 ; Strickland \& Stevens 2000; Fragile \etal 2004; Cooper \etal 2008, Fujita \etal 2009). However the detailed
physics of a radiatively driven galactic outflow is yet to be studied with a
simulation. 
In this work, we study the dynamics of an irradiated gas above an axisymmetric disc galaxy
by using hydrodynamical simulation.
Recently Hopkins \etal (2011) have explored the relative roles of radiation
and supernovae heating in galactic 
outflows, and studied the feedback on the star
formation history of the galaxy. Our goal here is different in the sense that we focus
on the structure and dynamics, particularly the effect of rotation, of the
wind. In order to disentangle the effects of various processes involved,
we intentionally keep the physical model simple. For example, we begin
with a constant density and surface brightness disk, then study the effect
of a radial density and radiation profile, and finally introduce rotation
of the disk, in order to understand the effect of each detail separately,
instead of performing one single simulation with many details put together.

\section{Gravitational and radiation fields}
The main driving force is radiation force and the 
containing force is due to gravity. We take the system to be composed of three components disc, bulge \& dark matter halo. We describe the forces due to these three constituents below. We take a thin galactic disc and a spherical bulge.  All these forces are given in cylindrical coordinates because we solve the fluid equations in cylindrical geometry.
  
\subsection{Gravitational field from the disc}
Consider a thin axisymmetric disc in $r\phi$ plane with surface mass density $\Sigma(r)$.
As derived in the Appendix, the
vertical and radial components of gravity due to the disc 
material at a point $Q$ above the disc with 
coordinates $(r, 0, z)$, are given by
\begin{eqnarray}
f_{disc,z} &=& \int_{\pp} \int_{\rp} d\pp \, d\rp \, \frac{z G \Sigma(\rp)\ \rp }{[r^2+z^2+\rp^2-2r\rp cos\pp]^{3/2}} \nonumber\\
f_{disc,r} &=& 
\int_{\pp}\int_{\rp} \, d\pp  \,d\rp
\frac{(r-\rp cos\pp)\  G \Sigma(\rp)\ \rp }{[r^2+z^2+\rp^2-2r\rp cos\pp]^
{3/2}} \
\label{eq:gravdisk}
\end{eqnarray}
The azimuthal coordinate of $Q$ is taken to be zero, because of axisymmetry.
The integration limit for $\pp=0$ to $2\pi$.

We consider two types of disc in our simulations, one with uniform surface
mass density and radius $r_d$ (UD), and another with an exponential distribution of surface
mass density (ED) with a scale radius $r_s$.
The surface mass density of
uniform surface density disc (\ie UD) is
\bea
\Sigma=\Sigma_0=\mbox{constant}
\label{eq:unid}
\eea
and in the case of a disc with exponentially falling density distribution (ED) 
\bea
\Sigma={\tilde{\Sigma}}_0\mbox{exp}(-\rp/r_s), ~~ r_s \equiv \mbox{scale length}
\,.
\label{eq:expod}
\eea
In case of UD (eqn \ref{eq:unid}), the integration limit would
be
$\rp=0$ to $r_d$, while for ED (eqn \ref{eq:expod}), the limits of the integration
run from $\rp=0$ to $\infty$. 
Numerically this means, we integrate up to a large number, increasing which will 
not change
the gravitational field by any significant amount.
We have chosen the $\Sigma$s in such a way that the total disc mass remains
same for the UD or ED. Therefore,
\be
\tilde{\Sigma}_0=\frac{\Sigma_0}{2}\left(\frac{r_d}{r_s}\right)^2 \,. 
\label{eq:two_sigma}
\ee

In Figure 1, we plot the 
contours of gravitational field strength and its direction vectors due to a UD
(left panel), and that for the ED (right panel).
Interestingly, discs with same mass but different surface density distributions, produces different
gravitational fields. For the UD the gravitational field is not spherical and the gravitational acceleration
is maximum at the edge of the disc. On the other hand, the field due to ED is 
closer to spherical configuration with the maximum being closer to the centre of the disc and falling off outwards.

\begin{figure}
\centerline{
\epsfxsize=0.5\textwidth
\epsfbox{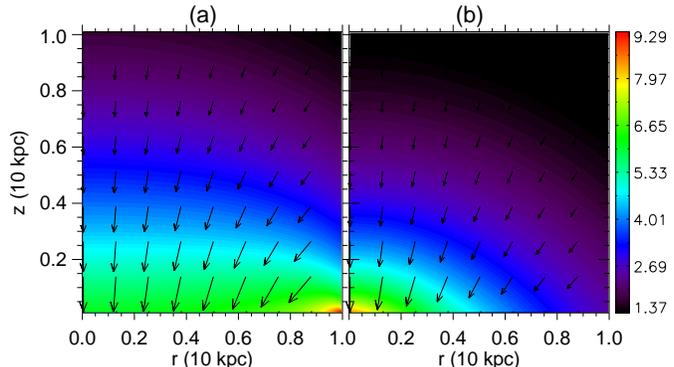}
}
{\vskip-3mm}
\caption{
Magnitude of gravitational force of the (a) uniform disc (UD)
(b) exponential disc (ED) in colours with
direction in arrows. Values are in the units of $G\Sigma_0 (= 4.5 
\times 10^{-9}$) dyne.
} 
\label{fig1:gunidisc}
\end{figure}

\subsection{Bulge and the dark matter halo}
We consider a bulge with a spherical mass distribution and constant density,
with mass $M_b$ and radius $r_b$. The radiation force due to the bulge is negligible as it mostly hosts the old stars.
 The gravitational force of the bulge is given by

\be
f_{bulge,r} = \left\{ \begin{array}{rl}
-\frac{G M_b r}{r_b^3} &\mbox{ if $R<r_b$} \\
\\
-\frac{G M_b r}{R^3} &\mbox{ otherwise}
\end{array} \right.
\ee
\be
f_{bulge,z} = \left\{ \begin{array}{rl}
-\frac{G M_b z}{r_b^3}\,, &\mbox{ if $R<r_b$} \\
\\
-\frac{G M_b z}{R^3} \,, &\mbox{ otherwise}
\end{array} \right.
\ee
where R = $\sqrt{r^2+z^2}$. 

We consider a NFW halo with a scaling with disc mass as given by Mo, Mao and White 
(1998; hereafter referred to as MMW98) where the total halo mass is $\sim 20$ times the total disc mass. The mass of 
an NFW halo has the following functional dependence on R 
\bea
M(R)= 4 \pi \rho_{crit} \delta_{0} R_s^3 \left[ \ln{(1+cx)}-\frac{cx}{1+cx}\right] \,
\eea
where
$x = \frac{R}{R_{200}}\,, c = \frac{R_{200}}{R_s} \,,
\delta_0 = \frac{200}{3}\frac{c^3}{ln(1+c)-c/(1+c)}$.
Here $\rho_{crit}$ is the critical density of the universe  at present
epoch, R$_s$ is scale radius of 
NFW halo and R$_{200}$ is the limiting radius of virialized halo within which the 
average density is 200$\rho_{crit}$.  
This mass distribution corresponds to the following potential,
\bea
\Phi_{NFW} = -4 \pi \rho_{crit} \delta_{0} R_s^3 \Bigl [ \ln{(1+R/R_s)} / R
\Bigr ]
\eea
The gravitational force due to the dark matter halo is therefore given by,
\begin{eqnarray}
f_{halo,r} = -\frac{\pd \Phi_{NFW}}{\pd r} = -\frac{r\ G M(R)}{(r^2+z^2)^{3/2}} ; \nonumber\\
f_{halo,z} = -\frac{\pd \Phi_{NFW}}{\pd z} = -\frac{z\ G M(R)}{(r^2+z^2)^{3/2}}
\,.
\end{eqnarray}
The net gravitational acceleration is therefore given by
\bea
& & F_{grav,r}=f_{disc,r}+f_{bulge,r}+f_{halo,r}=G\Sigma_0 f_{g,r}(r,z) \\ \nonumber
& & F_{grav,z}=f_{disc,z}+f_{bulge,z}+f_{halo,z}=G\Sigma_0 f_{g,z}(r,z)
\,.
\eea

\begin{figure}
\centerline{
\epsfxsize=0.5\textwidth
\epsfbox{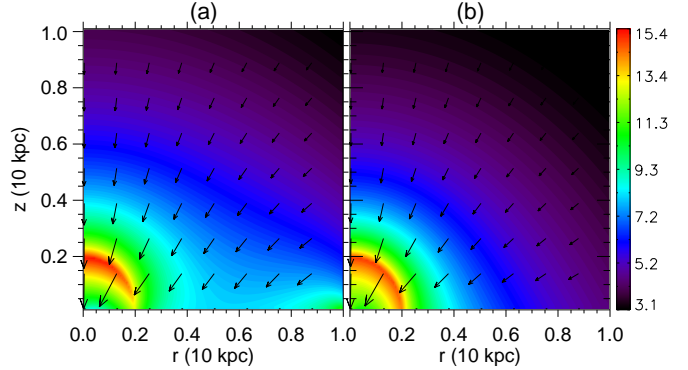}
}
{\vskip-3mm}
\caption{
Total gravitational force of the (a)
uniform disc (b) exponential disc in colors with
 direction in arrows. 
The values are in the same units as in Figure \ref{fig1:gunidisc}.
} 
\label{fig4:gtotexp}
\end{figure}

The gravitational field for both bulge and halo is spherical in nature,
although, that due to the bulge maximises at $r_b$.
However, the net gravitational field will depend on the relative strength of the three components.
In Figure \ref{fig4:gtotexp} (left panel), we plot the contours of total gravitational field strength
due to the bulge, the halo and an UD. The non-spherical nature of the gravitational field is evident.
A more interesting feature appears due to the bulge gravity. The net gravitational intensity maximizes
in a spherical shell of radius $r_b (=0.2 L_{ref}$; see section \S 3.1). Therefore, there is a possibility of piling up of outflowing 
matter at around a height $z\sim r_b$ near the axis. 
In the right panel of Figure (\ref{fig4:gtotexp}), we present the contours of 
net gravitational field due to an embedded exponential disc within a halo and a bulge.

\subsection{Radiation from disc and the Eddington factor}
We treat the force due to radiation pressure as it interacts with charged dust 
particles 
that are assumed to be strongly coupled to gas by Coulomb interactions 
and which drags the gas with it. The strength of
the interaction is parameterized by the dust opacity $\kappa$ which has the units
cm$^2$ gm$^{-1}$. 

Gravitational pull on the field point $Q(R,Z)$ due to the disc point $P(\rp,\pp,0)$ is along the
direction ${\overrightarrow{QP}}$ (see appendix). The difference in computing the radiation force arises due to the fact that one needs to account for the projection of the 
intensity at $Q$ (for radiation force from more complicated disc, see Chattopadhyay 2005).
For a disc with surface brightness $I(r)$,
we can find the radiation force by replacing $G\Sigma (r \prime)$ in 
eqn \ref{eq:gravdisk} by $I(\rp)\kappa/c$, and take into account the projection factor $z / \sqrt{r^2 +z^2 +r \prime ^2 -2 r r\prime \cos \pp}$.
Similar to the disc gravity, the net radiation force ${\overrightarrow{F}_{rad}}$ at any point will have the radial component
($F_{rad,r}$) and the axial component ($F_{rad,z}$) and are given by,
\begin{eqnarray}
F_{rad,r}(r,z) &=& \frac{\kappa z}{c}\int \int \frac{d\pp d\rp I(\rp)(r-\rp cos\pp)\ \rp}{[r^2+z^2+\rp^2-2r\rp cos\pp]^2} \\ \nonumber
&=& \frac{\kappa I_0}{c} f_{r,r}(r,z)
\end{eqnarray}
\begin{eqnarray}
F_{rad,z}(r,z) &=& \frac{\kappa z^2}{c}\int \int \frac{d\pp d\rp I(\rp)\rp}{[r^2+z^2+\rp^2-2r\rp cos\pp]^2} \\ \nonumber
&=& \frac{\kappa I_0}{c} f_{r,z}(r,z)
\end{eqnarray}
Since we have two models for disc gravity, we also consider two forms of disc 
surface brightness.
\bea
I=I_0=\mbox{constant, for UD} 
\label{eq:iunid}
\eea
and 
\bea
I={\tilde I}_0\mbox{exp}(-\rp/r_s) \,, \mbox{for ED}
\label{eq:iexpod}
\eea
If the two disc types are to be compared for identical luminosity, then one 
finds
\be
{\tilde I}_0=\frac{I_0}{2}\left(\frac{r_d}{r_s}\right)^2 \,.
\ee

 \begin{figure}
\centerline{
\epsfxsize=0.5\textwidth
\epsfbox{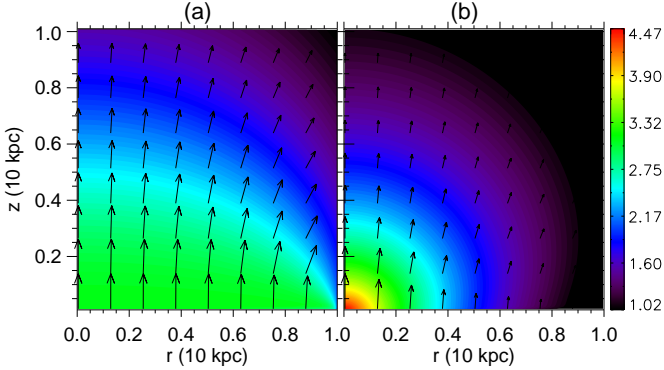}
}
{\vskip-3mm}
\caption{
Magnitude of force due to radiation from the (a) uniform disc, (b) exponential disc  for $\Gamma_0 =
 0.5$, with arrows for direction.
}
\label{fig5:raduni} 
\end{figure} 

The disc Eddington factor is defined as the ratio of the radiation force and 
the 
gravitational force (MQT05).
In spherical geometry this factor is generally constant at each point because both 
gravity and radiation has an inverse square dependence on distance. Although in the 
case of a disc, the two forces have different behaviour, we can still define an
Eddington parameter as 
$\Gamma = \frac{F_{rad}}{F_{grav}}$. In this case 
this parameter depends on the coordinates $r,\phi, z$ of the position under 
consideration. We can however define a parameter whose value is the Eddington factor
at the centre of the disc, i.e.,
\bea
\Gamma_0 = \frac{\kappa I}{2 c G \Sigma}.
\label{eq:gam0rel}
\eea
If $\Gamma_0 = 1$, then the radiation and gravity of the disc will cancel each other 
at the centre of the disc. We will parameterize our results in terms of $\Gamma_0$.
Therefore, the components of the net external force due to gravity and radiation is given by
\bea
& & {\cal R}_r=F_{grav,r}-F_{rad,r}=G\Sigma_0\left(f_{g,r}-2\Gamma_0f_{r,r}\right) \\ \nonumber
& & {\cal R}_z=F_{grav,z}-F_{rad,z}=G\Sigma_0\left(f_{g,z}-2\Gamma_0f_{r,z}\right)
\label{eq:forc}
\eea

In Figure \ref{fig5:raduni}, we plot the contours of radiative acceleration from an UD, and the same from an ED.
There is a significant difference between the radiation field above an ED and that above an UD.
While the radiation field from an UD is largely  vertical for small radii, but starts to diverge at the disc edge,
at $r\sim r_d$. One can therefore expect that for high enough $I$, the wind trajectory will diverge.
In case of ED, the radiation field above the inner portion of the disc  is strong
and decreases rapidly towards the outer disc.

\section{Numerical Method}
The hydrodynamic equations have been solved in this paper by using the TVD (\ie Total Variation Diminishing) code, which has been 
quite exhaustively used in cosmological and accretion disc simulations (see, Ryu \etal 1993, Kang \etal 1994, Ryu \etal 1995,
Molteni \etal 1996) and is based on a scheme originally developed by Harten (1983). We have solved the equations
in cylindrical geometry in view of the axial symmetry of the problem. This code is based on an explicit, second order accurate
scheme, and is obtained by first modifying the flux function and then applying a non-oscillatory first order accurate scheme to 
obtain a resulting 
second order accuracy (see, Harten 1983 and Ryu \etal 1993 for details). 

The equations of motion which are being solved numerically in the non-dimensional form is given by

\bea
\frac{\partial {\bf q}}{\partial t}+\frac{1}{r}\frac{\partial (r{\bf F}_1)}{\partial r}
+\frac{\partial {\bf F}_2}{\partial r}+\frac{\partial {\bf G}}{\partial z}={\bf S}
\label{eq:eqmo}
\eea
where, the state vector is
\bea
{\bf q}=\left(\matrix{\rho \cr \rho~v_r \cr \rho~v_{\phi} \cr \rho~v_z \cr E}\right),
\label{eq:varb}
\eea
and the fluxes are
\bea
{\bf F}_1=\left(\matrix{\rho~v_r \cr \rho~v^2_r \cr \rho v_r v_{\phi} \cr
\rho v_z v_r \cr (E+p)v_r}\right), ~~ {\bf F}_2=\left(\matrix{0 \cr p \cr 0 \cr 0 \cr 0}\right),
~~ {\bf G}= \left(\matrix{\rho v_z \cr \rho v_r v_z \cr \rho v_{\phi} v_z \cr
\rho v^2_z+p \cr (E+p)v_z} \right)
\label{eq:flux}
\eea
and the source function is given by
\bea
{\bf S}=\left[\matrix{0 \cr \frac{\rho v^2_{\phi}}{r}-\rho {\cal R}_r \cr -\frac{\rho v_r v_{\phi}}{r}
\cr -\rho {\cal R}_z \cr -\rho [v_r{\cal R}_r+v_z{\cal R}_z)}\right]
\label{eq:sorc}
\eea

\subsection{Initial and boundary conditions}
We do not include the disc in our simulations and only consider the effect
of disc radiation and total gravity on the gas being injected from the disc.
We choose the disc mass to be $M_d=10^{11}$ M$_{\odot}$ and assume it to be the unit of mass (\ie  $M_{ref}$). The unit of length (\ie $L_{ref}$)
and velocity (\ie  $v_{ref}$) are $r_d=10$ kpc and $v_c=200$ km s$^{-1}$, respectively. Therefore,
the unit of time is $t_{ref}=48.8$Myr. We introduce a normalization parameter $\xi$
such that $GM_d/v_c^2=\xi r_d$, which turns out to be $\xi=1.08$. Hence the unit of density
is $\rho_{ref}=6.77{\times}10^{-24}$g cm$^{-3}$ ($\sim 4 m_p$ cm$^{-3}$). 
All the flow variables have been made non-dimensional
by the choice of unit system mentioned above.

It is important to choose an appropriate
initial condition to study the relevant physical
phenomenon. 
We note that previous simulations of galactic outflows have considered a variety
of gravitational potential and initial ISM configurations. For example,
Cooper \etal (2008) considered the potential
of a spherical stellar bulge and an analytical expression for disc potential,
but no dark matter halo, and an ISM that is stratified in $z$-direction with
an effective sound speed that is $\sim 5$ times the normal gas sound speed.
Suchkov \etal (1994) considered the potential
of a spherical bulge and a dark matter halo and an initial ISM that is spherically
stratified. Fragile \etal (2004) considered a spherical halo and a
$z$-stratified ISM.
However,
in a recent simulation of outflows driven by supernovae from disc galaxies, Dubois
\& Teyssier (2008) found that the outflowing gas has to contend with infalling
material from halo, which inhibits the outflow for a few Gyr. Fujita \etal (2004)
also studied  outflows from pre-formed disc galaxies in the presence of a cosmological 
infall of matter. 

We choose a $z$-stratified gas
to fill the simulation box, with a scale height of $100$ pc. For the M$_2$ and
M$_3$ case (of exponential disc), we also assume a radial profile for the initial
gas, with a scale length of $5$ kpc. For the M$_3$ case, 
we further assume this
gas to rotate with $v_\phi$ decreasing with a scale height of $5$ kpc. These
values are consistent with the observations of Dickey \& Lockman (1990) 
and Savage
\etal (1997) for the warm neutral gas ($T \sim 10^4$ K) in Milky Way. 
We note that although the scale height for the 
warm neutral 
gas in our Galaxy is $\sim 400$ pc at the solar vicinity, this is expected to 
be smaller in the central region because of strong gravity due to bulge.   
The density of the gas just above the disc is assumed to be $0.1$ particles /cc
($0.025$ in simulation units). 

Furthermore, the adiabatic index of the gas is $5/3$ and the
gas is assumed initially to be at the same temperature corresponding
to an initial sound speed $c_s(ini)=0.1v_{ref}$, a value which is 
consistent with the values in our Galaxy
for the warm ionized gas with sound speed $\sim 18$ km s$^{-1}$.

\begin{figure}
\centerline{
\epsfxsize=0.5\textwidth
\epsfbox{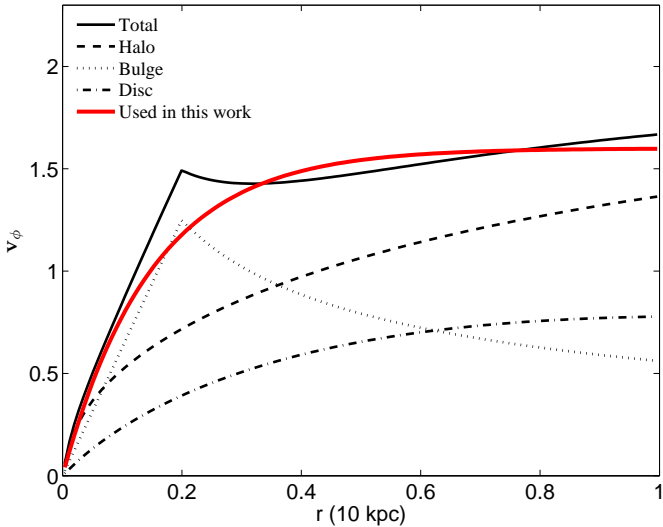}
}
{\vskip-3mm}
\caption{
Rotation curves corresponding to the gravitational fields of an exponential
 disc, bulge and halo
are shown here in the units of $v_{ref} [=200$ km s$^{-1}]$, along with the 
total rotation curve. The approximation used in 
our simulation is shown by thick red line.
}
\label{allrot}
\end{figure}

Our computation domain is $r_d~\times~r_d$ in the $r-z$ plane, with a resolution 
$512~\times~512$ cells. The size of individual computational cell is $\sim 20$ pc. 
We have imposed reflective boundary condition around the axis and zero rotational velocity on the axis.
Continuous boundary conditions are imposed at $r=r_d$ and $z=r_d$. 
The lower boundary is slightly above the
galactic disc with an offset $z_0=0.01$. 
We impose fixed boundary condition at lower $z$ boundary. 
The velocity of the injected matter is $v_z(r,z_0)=v_0=10^{-5} v_{ref}$,
and its density is given by,
\bea
\rho(r,z_0) &=& \rho_{z_0}, ~~ \mbox{for UD} \\ \nonumber
 &=& \rho_{z_0}\mbox{exp}\left(-\frac{r}{r_s}\right), ~~ \mbox{for ED} \,.
\label{eq:den}
\eea
The density of the injected matter at the base $\rho_{z_0}=0.025$ 
(corresponding to $0.1$ protons per cc). 

For the case of exponential disc with
rotation (M$_3$), we assume for the injected matter to have an angular momentum
corresponding to an equilibrium rotation profile. We show in Figure
\ref{allrot} the rotation curves at $z=0$ for all components (disc, bulge
and halo) separately and the total rotation curve. We use the following
approximation (shown by thick red line in Figure \ref{allrot}) which matches
the total rotation curve,
\be
v_{\phi}(r,z_0) = 
1.6 \, v_c~[1-\mbox{exp}(-r/0.15r_d)] \,. 
\label{eq:vrot}
\ee

We assume a bulge of mass $M_b=0.1 M_{ref}$ and radius $r_b=0.2 
L_{ref}$. The scale radius for NFW halo (R$_s$) is determined for a
halo mass $M_h=20 M_d$, as prescribed by MMW98. The
corresponding disc scale radius is found to be $r_s\sim 5.8$ kpc, again
using MMW98 prescriptions. Therefore we set the disc scale length for
the ED case to be $r_s\sim 0.58 L_{ref}$.

The above initial conditions have been chosen to satisfy the following
requirements in order to sustain a radiatively driven wind as simulated here.
\begin{enumerate}
\item  The strong coupling between dust grains and gas particles require
that there are of order $\sim m_d/m_p$ number of collisions between protons
and dust
grains of mass $m_d \sim 10^{-14}$ g, for size $a \sim 0.1 \, \mu$m  with
density $\sim 3 $g cm$^{-3}$. To ensure sufficient number of collisions,
the number density of gas particles should be $n \ge {m_d \over m_p} {1 \over 
\pi a^2} {1 \over L_{ref}} \sim 10^{-3} $ cm$^{-3}$, for $L_{ref}=10$ kpc.

\item  The time scale for radiative cooling  of the gas, assumed to be
at $T\sim 10^4$ K, is $t_{cool}\sim {1.5 kT \over n \Lambda}$, 
where  $\Lambda \sim 10^{-23}$ erg cm$^3$ s$^{-1}$
(Sutherland \& Dopita 1993; Table 6) for solar metallicity. The typical density
filling up the wind cone
in the realistic case (M$_3$) is $\sim 10^{-3}\hbox{--}10^{-4}$ cm$^{-3}$, which gives $t_{cool}\sim 8\hbox{--}80$ Myr
and the dynamical time scale of the wind is  $t_{ref}\sim 50$ Myr. Hence
radiative cooling is marginally important and we will address the issue of radiative cooling in a future paper.

\item  Radiative transfer effects are negligible since the total opacity
along a vertical column of length $L_{ref}$ is $\kappa (n m_p) L_{ref}
\sim 0.003$, for $n\sim 10^{-3}$ cm$^{-3}$ and $\kappa \sim 100$ cm$^2$ g$^{-1}$.

\item  The mediation of the radiation force by dust grains also implies that the
gas cannot be too hot for the dust grains to be sputtered. The sputtering
radius of grains embedded in even in a hot gas of temperature  T$\sim 10^5$ K
is $\sim 0.05 (n/0.1 \, / {\rm cc}) \, \mu$m in a time scale of $100$ Myr
(Tielens \etal 1994), and this effect is not important for the temperature and density
considered here.

\end{enumerate}

\begin{table}
\begin{center}
 \caption{Models.}
\label{modtable}
\vskip 0.2cm
  \begin{tabular}{|c|c|c|c|}
                  \hline
{\small Model name}&$\Gamma_0$&$v_{\phi}$&Disc type \\
                   \hline
                   \hline
M$_1$&$2.0$&$0.0$& UD \\
                   \hline
M$_{2}$&$2.0$&$0.0$& ED\\
  \hline
M$_{3}$&$2.0$&$1.0$& ED \\
                   \hline
\end{tabular}      
\end{center}
\end{table}

\subsection{Simulation set up}
We present 3 models with parameters listed in the  Table \ref{modtable}. The initial condition for all the models are described in \S 3.1.
The boundary condition is essentially same, except that the mass flux into the computational domain
from the lower $z$ boundary depends on the type of disc. As has been mentioned in section 3.1,  we keep the velocity of injected 
matter very low, $v_z(r,z_0)=v_z(ini)=10^{-5} v_{ref}$, so that it does not affect the dynamics. 
The three models have been constructed
by a combination of different values of three parameters $\Gamma_0$, $v_{\phi}$ and the distribution
of the density in the disc.
Model M$_3$ has been run for different values of $\Gamma_0$, to ascertain the effect of radiation.

\begin{figure}
\centerline{
\epsfxsize=0.5\textwidth
\epsfbox{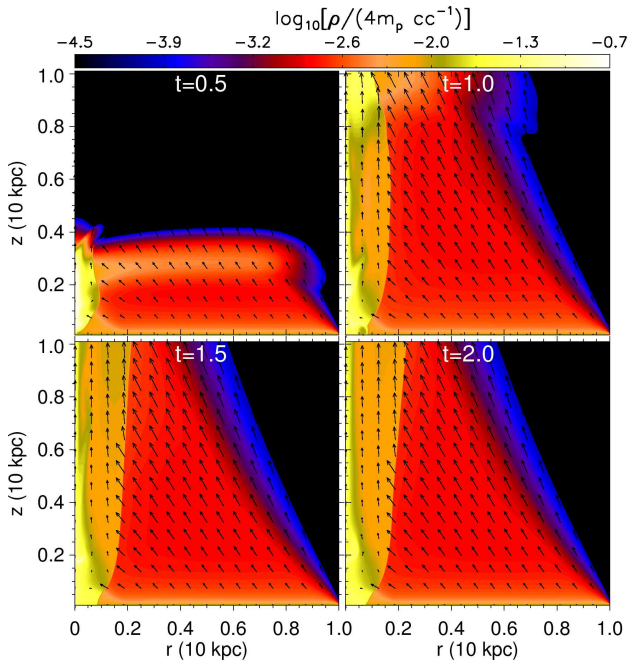}
}
{\vskip-3mm}
\caption{
M$_1:$ Logarithmic density contours for radiation driven wind from UD for four snapshots running
up to $t=98$ Myr, with velocity vectors shown with arrows.
Densities are colour-coded according to the computational unit of density, $6.7
\times 10^{-24}$ g cm$^{-3} \sim 4 m_p$ cm$^{-3}$.
}
\label{fig8:m1}
\end{figure}
\section{Results}
In Figure \ref{fig8:m1}, we present the model $M_1$ for a constant surface
density disc (UD).
The density contour and the velocity vectors for the wind
are shown in four snapshots in Figure (\ref{fig8:m1})  upto a time $t=98$
Myr (corresponding to $t=2$ in computational time units).
There are a few aspects of the gaseous flow that we should note here. Firstly,
the disc and the outflowing gas in this case has no rotation ($v_\phi=0$). In
the absence of the centrifugal force due to rotation which might have reduced the
radial gravitational force, there is a net radial force driving the gas inward.
At the same time, the radiation force, here characterized by $\Gamma_0=2$, propels
the gas upward (the radial component of radiation being weak). 
The net result after a few Myr is that the gas in the region
near the pole moves in the positive $z$ direction, and there is a density
enhancement inside a cone around the pole, away from which the density and
velocities decrease. 

\begin{figure}
\centerline{
\epsfxsize=0.5\textwidth
\epsfbox{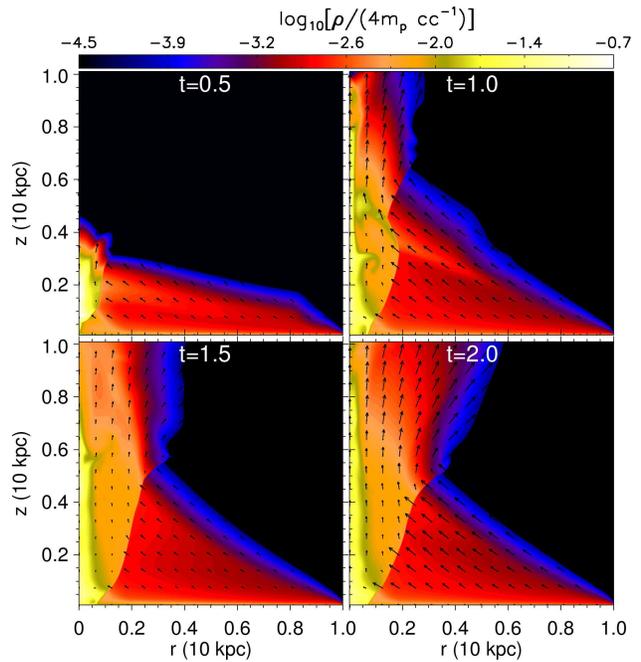}
}
{\vskip-3mm}
\caption{
M$_2:$ Logarithmic density contours for radiation driven wind from ED for four snapshots running
up to $t=98$ Myr, with velocity vectors shown with arrows.
}
\label{figm2}
\end{figure}

Also, because of the strong gravity of the bulge,
the gas tends to get trapped inside the bulge region, and even the gas at larger 
$r$ tends to get dragged towards the axis. This region puffs due to accumulation of
 matter. Ultimately the
radiative force drives matter outwards in the form of a plume.

Next, we change the disc mass distribution and simulate
the case of wind driven out of an exponential disc (ED). We show the results
in Figure \ref{figm2}. Since both gravity and radiation forces in this case
of exponential disc are quasi-spherical in nature, therefore in the final snapshot
the flow appears to follow almost radial streamlines. Although in the vicinity
of the disc, the injected matter still falls towards the axis, but this is not
seen at large height as was seen in the previous case of M$_1$. This makes the
wind cone of rising gas more diverging than in the case of UD (M$_1$).

\subsection{Rotating wind from exponential disc}
\begin{figure*}
\centerline{
\epsfxsize=0.7\textwidth
\epsfbox{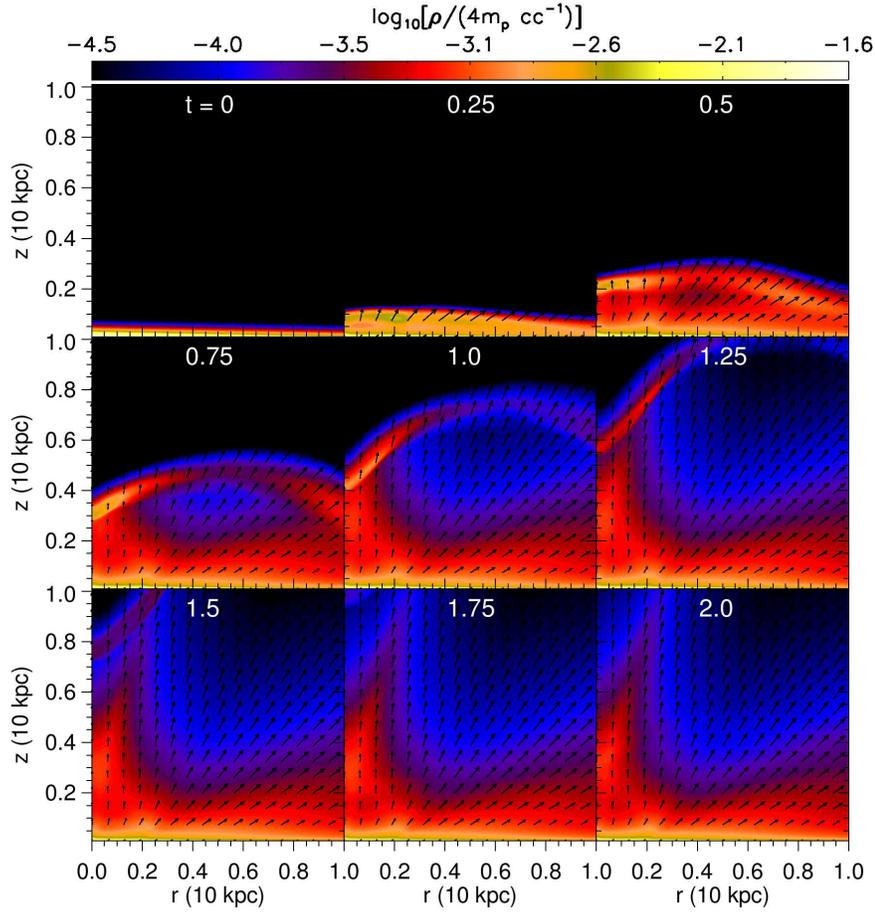}
}
{\vskip-3mm}
\caption{
M$_3$: Contours of $\log _{10} (\rho)$ and ${\bf v}$-field of radiation driven wind with $\Gamma_0=2.0$ from an
ED. t = 2 corresponds to 98 Myr.
} 
\label{figm3}
\end{figure*}

The direction of the fluid flow in M$_1$ and M$_2$ is by and large
towards the axis, and this flow is mitigated in the presence of rotation in 
the disc and injected gas. 
In the next model  M$_3$, we consider rotating  matter being
 injected into the
computational domain and which follows a $v_{\phi}$ distribution given by Eq. 
(\ref{eq:vrot}).
This is reasonable to assume since the disc from which the wind is supposed to blow,
is itself rotating. In M$_3$, we simulate rotating gas being injected above a ED 
and being driven by a radiation force of $\Gamma_0=2$.
We present nine snapshots of the M$_3$ case in Figure \ref{figm3}.

The first six snapshots of Figure \ref{figm3}
show the essential dynamics of the outflowing gas.
The fast rotating matter from the outer disc is driven outward because
the radial gravity component is balanced by rotation. Near the central
region, rotation is small and also the radial force components are small.
Therefore the gas is mostly driven vertically.
The injected gas reaches a vertical height of $\sim 5$ kpc in a time
scale of $\sim 37$ Myr (t=0.75). The flow reaches a steady state after $\sim 60$ Myr (t=1.25). 
In the steady state we find
a rotating and mildly divergent wind.

\begin{figure}
\centerline{
\epsfxsize=0.5\textwidth
\epsfbox{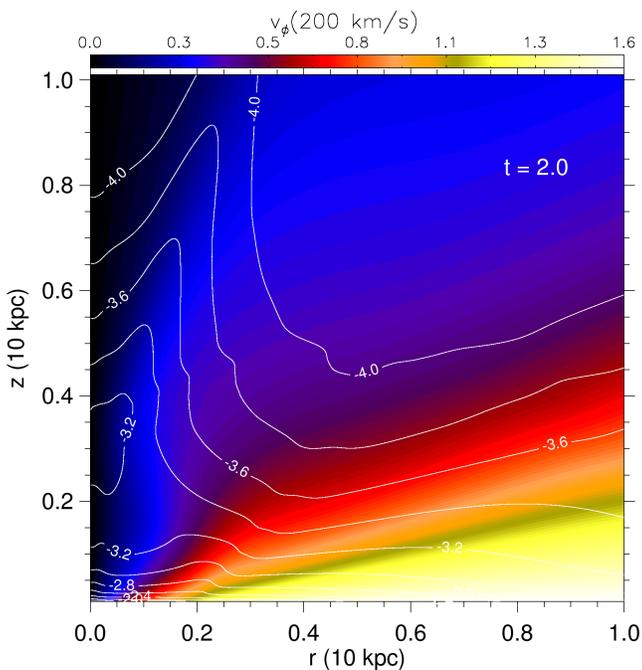}
}
{\vskip-3mm}
\caption{
The rotation velocity $v_\phi$  for the case M$_3$ at a time of 98 Myr
is shown in colours.   
Contour lines of log$_{10}$$(\rho)$ are plotted over it.  
}
\label{vphii}
\end{figure}

We show the azimuthal velocity contours in Figure \ref{vphii} in colour
for the fully developed wind (last snapshot in M$_3$), and superpose on it the 
contour lines of $\rho$. The density contours clearly show
a conical structure for outflowing gas. The rotation speed of the gas
peaks at the periphery of the cone, and is of order $\sim 50\hbox{--}100$ km
s$^{-1}$.
Compared to the disc rotation speed, the rotation speed of the wind region
is somewhat smaller. In other words, we find the wind 
mostly consisting of low-angular momentum gas lifted from the disc.

\begin{figure}
\centerline{
\epsfxsize=0.5\textwidth
\epsfbox{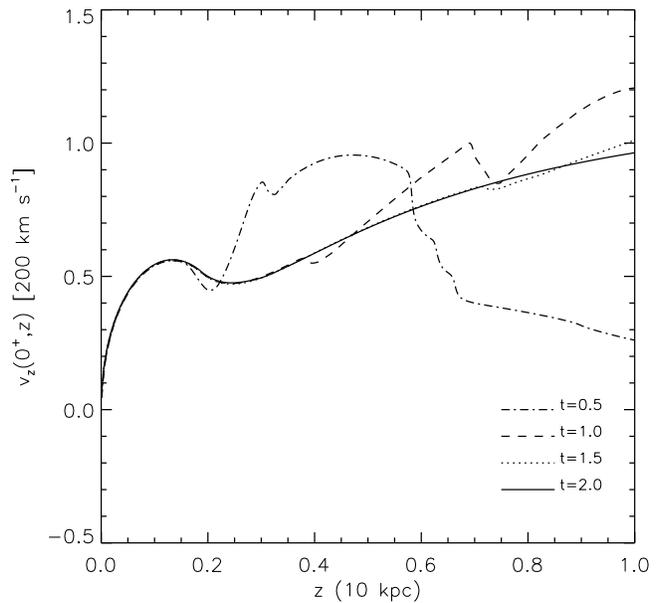}
}
{\vskip-3.0mm}
\caption{
The axial velocity $v_z(0^+,z)$ with $z$  at different time steps for the model M$_3$. t = 2.0 corresponds to a time of $98$Myr.
}
\label{fig9:vzm2-1}
\end{figure}
\begin{figure}
\centerline{
\epsfxsize=0.5\textwidth
\epsfbox{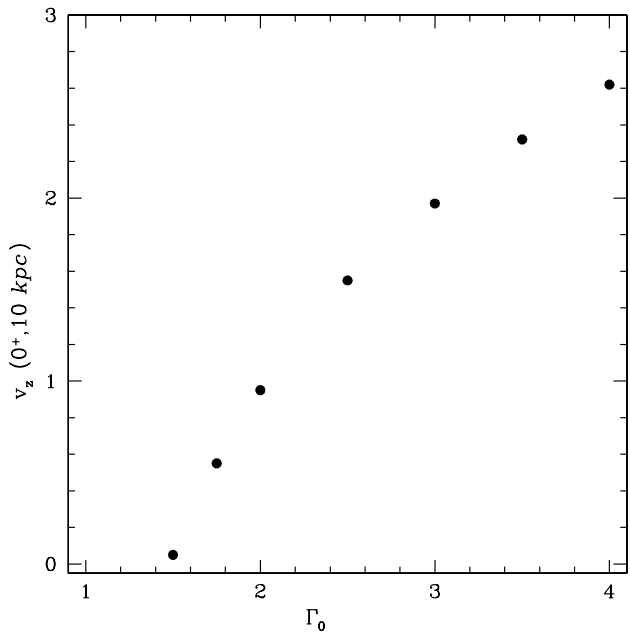}
}
{\vskip-3mm}
\caption{
The axial velocity $v_z(0^+,10 {\it kpc})$ in simulation units $v_{ref} = 200$ km s$^{-1}$ with $\Gamma_0$,
at a time $t\sim10^2$ Myr.
}
\label{fig9:last}
\end{figure}

We plot the velocity of gas close to the axis in Figure \ref{fig9:vzm2-1} for
different times in this model (M$_3$), using
${\bf v}(0,z)\sim v_z(0^+,z)$.  The velocity profile in the snapshots at
 earlier time
fluctuates at different height, but becomes steady after t $\ge 1.5$, as does
the density profile.

We have run this particular case of ED with rotation (model M$_3$) 
for  different values of $\Gamma_0$. In order to illustrate the results of these
runs, we plot  
the  $z$-component of velocity ($v_z (0^+,10\ kpc)$) at 10 kpc and at  simulation time, 
$t=2$ as a 
function of $\Gamma_0$ in Figure \ref{fig9:last}.
We find that significant wind 
velocities are obtained for $\Gamma_0 \gtrsim 1.5$ and 
wind velocities appear to 
rise linearly with $\Gamma_0$ after this critical value is acheived. 
Sharma \etal (2011) found this critical value to be $\Gamma_0 \sim 2$ for
a constant density disc and wind launched above the bulge. For the
realistic case of an exponential disc, we find in the present simulation
the critical value to be somewhat smaller than but close to the analytical
result. The important point is that the critical $\Gamma_0$ is not unity. This is because
the parameter $\Gamma_0$ is not
a true Eddington parameter since it is defined in terms of disc gravity
and radiation, whereas halo and bulge also contribute to gravity.

\section{Discussions}
Our simulation differs from earlier works (e.g. Suchkov et.al. 1994) mainly in that we
specifically target warm outflows and the driving force is radiation pressure. 
Most of the previous simulations
of galactic wind have used energy injected  from supernovae blasts as a 
driving force. 
However, with the ideas presented in Murray \etal (2005), which worked
out the case of radiation pressure in a spherical symmetric set-up, it beomes
important to study the physics of this model in an axisymmetric set up, as
has been done analytically by Sharma \etal (2011) (see also, Zhang \& Thompson
2010). 
Also we have tried to capture all features of a typical disc galaxy like a bulge and a dark matter 
halo, and a rotating disc. Recent analytical works (Sharma \& Nath 2011) and simulations (Hopkins \etal 2011) have shown 
that outflows from massive galaxies ($M_{halo}\ge 10^{12}$ M$_{\odot}$)
have different characteristics than those from low mass galaxies. Outflows from massive galaxies
are mostly driven by radiation pressure and the fraction of cold gas in the halos of massive
galaxies is large (van de Voort \& Schaye 2011). Our simulations presented here addresses these outflows in
particular.

We have parameterized our simulation runs with the disc Eddington factor $\Gamma_0$, and it is
important to know the corresponding luminosity for a typical disc galaxy, or the equivalent 
star formation rate. 
For a typical opacity of a dust and gas mixture ($\kappa \sim 200$ cm$^2$ 
g$^{-1}$) (Draine 2011), the correspondig mass-to-light ratio requirement
for $\Gamma_0 \gtrsim 1.5$ is that $M/L \le 0.03$.
Sharma \etal (2011) showed that for the case of an instantaneous star formation,
$\Gamma_0 \gtrsim 2$ is possible for an initial period of $\sim 10$ Myr after the starburst. However
for a continuous star formation, which is more realistic for disc galaxies, Sharma \& Nath (2011)
found that only ultra luminous infrared galaxies (ULIGs), with star formation rate larger than
$\sim 100$ M$_{\odot}$ yr$^{-1}$ and which are also massive, 
are suitable candidates for such large values of $\Gamma_0$,
and for radiatively driven winds.

The results presented in the previous sections show that the outflowing gas 
within the central
region of a few kpc tends to stay close to the pole, and does not move outwards because of its low angular momentum. This  
 makes the outflow somewhat collimated.
Although outflows driven by SN heated hot wind also produces a conical structure (e.g., Fragile
\etal 2004) emanating from a breakout point of the SN remnants, there is  a qualitative difference
between this case and that of radiatively driven winds as presented in our simulations. While
it is the pressure of the hot gas that expands gradually as it comes out of a stratified
atmosphere, in the case of a radiation driven wind, it is the combination of mostly the lack of rotation 
and almost vertical radiation driving force in the central region that produce the collimation effect.

We also note that the conical structure of rotation in the outflowing gas is similar to 
the case of outflow in M82 (Greve 2004), where one observes a diverging and rotating periphery of
conical outflow.

We have not considered radiative cooling in our simulations, since for typical density in the wind
the radiative cooling time is shorter  or comparable 
than the dynamical time. However, there are regions of higher
density close to the base and radiative cooling can be important there. We will address this point
in a future paper.

From our results of the exponential and rotating disc model, we find the
wind comprising of low-angular momentum gas lifted from the disc.
It is interesting to 
note that recent simulations of supernovae driven winds have also claimed a similar result
(Governato \etal 2010). Such loss of low angular momentum gas from the disc may have important
implication for the formation and evolution of the bulge, since the bulge population is 
deficient in stars with low specific angular momentum. Binney, Gerhardt \& Silk (2001) have
speculated that outflows from disc that preferentially removes low angular momentum material
may resolve some discrepancies between observed properties of disc and results of numerical
simulations.

As a caveat, we should finally note that the scope and predictions of
our simulation is limited by the simple model of disc radiation adoped here.
In reality, radiation from disks is likely to be confined in the vicinity
of star clusters, and not spread throughout the disk as we have assumed here.
This is likely to increase the efficacy of radiation pressure, but which is
not possible within the scope of an axisymmetric simulation.

\section{Summary}
We have presented the results of hydrodynamical (Eulerian) simulations of radiation driven winds
from disc galaxies. After studying the cases of winds from a constant surface density
disc and exponential disc without rotation, we have studied a rotating outflow
originating from an exponential disc with rotation. We find that the outflow speed
increases rapidly with the disc Eddington parameter $\Gamma_0=\kappa I/(2 c G \Sigma)$ for $\Gamma_0 \ge 1.5$, consistent with theoretical expectations. The density
structure of the outflow has a conical appearance, and most of the ouflowing
gas consists of low angular momentum gas.

We thank Yuri Shchekinov for  constructive comments and critical reading of the manuscript.
IC acknowledges the hospitality of the Astronomy and Astrophysics Group of Raman Research Institute,
where the present work was conceived. DR was supported by National Research Foundation of Korea
through grant 2007-0093860.

\section{Appendix}
Consider a razor thin disc in r$\phi$ plane as illustrated in the fig. Now our task is to calculate the force components at any arbitrary point above the disc. Let us consider an annulus of the disc between $\rp$ and $\rp+d\rp$.  Area of the element at point P($\rp,\pp,0$) is $\rp d\rp d\pp$. Also take a field point Q(r,0,z) above the disc plane. Azimuthal coordinate of Q is taken to be zero for simplicity as we know that azimuthal force components are zero due to symmetry. Let QN and QM be perpendiculars from Q on the x and the z axis, respectively. So we can write,

\begin{eqnarray}
 PN^2 &=& (r-\rp cos\pp)^2 + (\rp sin\pp)^2 \nonumber \\
 PQ^2 &=& PN^2 + z^2 = r^2+z^2+\rp^2-2r\rp cos\pp \nonumber \\
sin\angle PQN &=& \frac{PN}{PQ} 
\end{eqnarray}
The  gravitational force due to the small area element at P is given by
\be
d{\bf F_g} = \frac{G\ dm\ PQ}{(PQ)^3} \hat n\ ;\\
 dm = \rp d\rp d\pp \Sigma(\rp)
\ee
Here $\Sigma(\rp)$ is the surface density of the disc.
Now the z component of this force is
\be
dF_{g,z} = |d{\bf F_g}| \frac{z}{PQ} = \frac{z G \Sigma(\rp)\ \rp d\rp d\pp}{[r^2+z^2+\rp^2-2r\rp cos\pp]^{3/2}}
\ee
To calculate the radial component, let PS be the perpendicular from P on the x-axis. Then, we have sin$\angle$SPN = SN/PN = (r-$\rp cos\pp$)/PN. 
Component of the force along the direction of PN is
\be
dF_{g,PN} = |d{\bf F_g}| sin\angle PQN = |d{\bf F_g}|\frac{SN}{PN}
\ee
So the radial component is
\begin{eqnarray}
dF_{g,r} &=& d F_{g,PN} sin\angle SPN = |d{\bf F_g}| \frac{SN}{PN}\frac{PN}{PQ} \nonumber\\
 &=& \frac{(r-\rp cos\pp)\  G \Sigma(\rp)\ \rp d\rp d\pp}{[r^2+z^2+\rp^2-2r\rp cos\pp]^{3/2}}
\end{eqnarray}

\begin{figure}
\centerline{
\epsfxsize=0.45\textwidth
\epsfbox{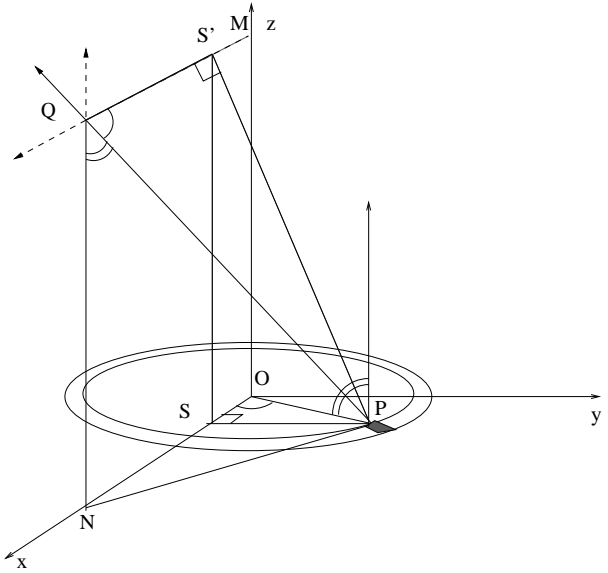}
}
{\vskip-3mm}
\caption{
Schematic diagram for the calculation of gravitational force due to disc
in the $xy$-plane.
We consider an annulus in the disc and
an element of area around the point P ($r \prime, \phi, 0)$ in this
annulus
is considered here in order to compute the force at a point Q ($r,0,z$)
whose azimuthal coordinate $\phi=0$. The point S ($r\prime \cos \phi,
0,0$) is the foot of the perpendicular drawn from P on the $x$-axis.
The point S$\prime$ ($r \prime,
0,z$) is at the intersection of the vertical from S (along $z$-axis)
and the line parallel to $x$-axis at height $z$.  The angle
$\angle SQS\prime = \cos ^{-1} \Bigl [ {S\prime Q \over PQ} \Bigr ]$,
and $\angle PQN =\cos ^{-1} \Bigl [ {QN \over PQ} \Bigr ]$.
}
\end{figure}

\end{document}